\documentclass[12pt]{article}
\usepackage{epsfig}
\usepackage{latexsym}
\begin{document}

\title{60 years of gauge fields.}
\author{ A.A.Slavnov \thanks{E-mail:$~~$ slavnov@mi.ras.ru}\\
V.A.Steklov Mathematical Institute\\ Gubkina street 8, GSP-1,119991,\\
and Moscow State University}

\maketitle

\begin{abstract}
the development of nonabelian gauge fields for last 60 years is reviewed. The new method of quantization of gauge fields applicable beyond perturbation theory is proposed.
\end{abstract}

\section{Introduction.}

Most people, presented here, hardly realize the situation which existed in high energy 
physics at 50-60- ties.

The Standard model and Weinberg- Salam model did not exist and even the question what
are the particle which transfer the interaction was not clear. 

Classical gauge fields, introduced by Yang and Mills (\cite{YM}), gave the answer to
this question.

It is postulated in the field theory that  any particle is described by the field, which for any $x$ 
belongs to representation of semi simple compact Lie group, that is if $\omega(x)$ is a function with values in the gauge group $\Omega$
\begin{equation}
\psi(x) \rightarrow \psi(x)^{\omega}=\Gamma[\omega(x)]\psi(x)
\label{1}
\end{equation}
The symbol $\Gamma$ denotes the matrix of the representation according to which the field $\psi$ transforms. 
The field configurations $\psi(x), A_{\mu}(x)$ and $\Gamma[\omega(x)]\psi(x), A_\mu^{\omega(x)}$ describe the same physical situation.
This principle is called the relativity principle in the charge space. It allows uniquely to introduce the interaction by changing in the free 
Lagrangian the usual derivatives by the covariant ones,
\begin{equation}
\partial_\mu \rightarrow D_\mu=\partial_\mu-\Gamma(A_\mu)
\label{2}
\end{equation}

In physical language $A_\mu(x)$ is called the gauge field. The function $A_\mu(x)$ may be presented as a sum $A_\mu(x)=A_\mu^a(x)T^a$,
where $T^a$ are the generators of the gauge group, which can be normalized as follows
\begin{equation}
tr(T^aT^b)=-2 \delta^{ab}
\label{2a}
\end{equation}
Then
\begin{equation} 
 [T^a,T^b]=t^{abc}T^c
\label{2b}
\end{equation}
where $t^{abc}$ are absolutely antisymmetric structure constants. 
It describes the parallel transport in the charge space. From the point of view of mathematics the Yang-Mills field  is 
the connection in the principle bundle. Under the gauge transformation it transforms as follows
\begin{equation}
A_\mu \rightarrow A_\mu^{\omega}=\omega (A_\mu) \omega^{-1}+\partial_\mu \omega \omega^{-1}
\label{3}
\end{equation} 

It is interesting to note that the notion of connection appeared in mathematics almost simultaneously with physics.
The first mathematician who introduced this notion was Lihnerovich ({\cite{L}).

The free Lagrangian for the Yang-Mills field is the square of the curvature tensor
\begin{equation} 
L=\frac{1}{8g^2}tr(F_{\mu \nu}F_{\mu \nu})
\label{4}
\end{equation}
\begin{equation}
F_{\mu \nu}=\partial_\mu A_\nu-\partial_\nu A_\mu+[A_\mu,A_\nu]
\label{5}
\end{equation}
This Lagrangian obviously describe the massless field. It prevented for a long time its using for description of weak interactions,
which are known to be short range, that is correspond to the exchange by the massive particle.

In the works by P.Higgs (\cite{Hi}, F.Englert and R.Broute (\cite{BE}) it was shown how one can 
introduce the mass term for the vector field preserving the gauge invariance of the model.
To do that one should introduce the gauge invariant interaction of the scalar field with a special potential, generating 
the spontaneous breaking of the global symmetry of the model, but preserving the gauge invariance.
The corresponding Lagrangian for the group $SU(2)$ looks as follows
\begin{equation}
L=(D_\mu \varphi(x))^*(D_\mu \varphi(x))+m^2(\varphi(x))^*(\varphi(x))-\lambda (\varphi(x)^*\varphi(x))^2-\frac{1}{4g^2}F^a_{\mu \nu}F^a_{\mu \nu}
\label{6}
\end{equation}

The potential
\begin{equation}
-V=m^2\varphi^*(x)\varphi(x)-\lambda(\varphi^*(x)\varphi(x))^2
\label{6a}
\end{equation}
where
\begin{equation}
\varphi^(x)=(\frac{iB_1+B_2}{\sqrt{2}}; \frac{\sigma-iB_3}{\sqrt{2}})
\label{6b}
\end{equation} 
is unstable and generates the spontaneous breaking of the goobal symmetry responsible for the conservation of the charge of the 
field $\varphi$. From the formal point of view the fields $\varphi$ acquire the imaginary mass. To develop the perturbation theory 
one should shift the fields $\sigma$ by the constant $\varphi \rightarrow \varphi+C$. Here $C$ is a constant doublet with zero first 
component and real second component.
Clearly the shift of the fields by constant cannot break the gauge invariance of the theory, but the form of transformations change.

Before the change of variables the gauge transformation of the fields $\varphi$ was a
phase transformation but after the change it includes the shift by arbitrary function, 
and $\varphi$ together with the field $A_\mu(x)$ plays the role of a gauge field. Under the shift of the field $\varphi$ by a constant
the term $(D_\mu\varphi)^*(D_\mu\varphi)$ generates the mass term for the vector field.

The main merit of H.Higgs and F.Englert and R.Brout is the fact that they showed how 
to introduce the vector field mass without breaking the gauge invariance.
One can notice that both papers appeared in 1964, and the correct quantization procedure for nonabelian gauge fields was 
constructed by L.Faddeev and V.Popov (\cite{FP}) and independently by B. DeWitt (\cite{DW}), three years later. For that reason strictly speaking
H.Higgs and F.Englert and R.Brout solved the problem only in Abelian case. For nonAbelian fields the corresponding procedure 
was constructed by T.Kibble (\cite{K}), following the same ideas.

The gauge invariant action describes the constrained system,  which does not allow 
to apply directly the methods based on the Hamiltonian character of the action.   
\begin{equation}
A=\int dx[p\dot{q}-H(p,q)+\lambda^a C^a]
\label{7}
\end{equation}
The first people who succeeded to construct 
the Lorentz invariant diagram technik for the Yang-Mills field were L.Faddeev and V.Popov and B.DeWitt. They reduced the 
constrained system, describing the gauge field, to the really Hamiltonian system, described by the Hamiltonian $H^*$ , 
solving the constraints and applying the
suitable gauge conditions. $H(p,q) \rightarrow H^*(p^*,q^*)$. These methods may be used as for the systems without 
spontaneous symmetry breaking and for the spontaneously broken systems (\cite{Ho}). But later the other methods of quantization were 
proposed. 

I have in mind the methods based on the BRST symmetry of the effective action (\cite{CF}), (\cite{KO}). These methods do not 
appeal for their justification to the non relativistic Coulomb gauge, and allow to work always without breaking the manifest
Lorentz invariance. They also are easily generalized to more complicated gauge invariant systems, such as strings. Moreover, as I am going 
to show these methods may be applied beyond perturbation theory as well. All the methods 
developed so far strictly speaking are applicable only in perturbation theory.

The construction of manifestly Lorentz invariant local formulation of QED is impossible without 
introduction of the vector-potential $A_\mu$, which describes not only physical excitations corresponding to three dimensionally
transversal components, but also the longitudinal and time-like photons. At the same time the action of electromagnetic field is gauge 
invariant. Such situation is typical for the field theory- one introduces the unphysical degrees of freedom, and simultaneously introduces
a new symmetry, which provides their decoupling. It is used for example in the BRST quantization of gauge systems.        

We shall use this idea for the construction of the quantization procedure of nonabelian gauge fields,
free of ambiguity beyond perturbation theory (\cite{Sl1}, \cite{Sl2},\cite{Sl3}, \cite{Sl4}, \cite{QS},\cite{QS1}).

The procedure of quantization of gauge fields includes the hypothesis, that if the  
gauge condition is imposed, the Cauchi problem for the gauge field becomes uniquely solvable. It is true for QED, based on the Abelian 
group $U(1)$. Choosing the Coulomb gauge $\partial_iA_i=0, \quad i=1,2,3$ we see that any gauge transformed field must satisfy
the equation
\begin{equation}
\partial_i A_i^\epsilon=0=\partial_i A_i+\triangle \epsilon=0
\label{8}
\end{equation}
The condition $\partial_i A_i=0$ is fulfilled therefore we must have $\triangle \epsilon=0$, that is 
$\epsilon$ is a harmonic function. The harmonic function has an extremum at the boundary, therefore $\epsilon=0$.

Let us pass to the nonabelian group $SU(N)$
In the Coulomb gauge the field $A_i$ must satisfy the equation $\partial_i A_i^\epsilon=0$.
For infinitesimal $\epsilon$ it follows
\begin{equation}
\triangle \epsilon_a+gt^{abc}A_i^b\partial_i \epsilon^c=0
\label{9}
\end{equation}
It was noticed by V.Gribov(\cite{Gr}), that this equation has nontrivial solutions vanishing at spatial infinity.
Therefore the uniqueness of the solution of the Cauchi problem is broken.
It does not influence the perturbation theory. In this case 
the eq.(\ref{9}) has only trivial solution.
If one ignores this problem the transition from one 
gauge to another becomes singular. The problem of quantization of nonabelian gauge fields beyond perturbation theory remained open.

The problem of Gribov ambiguity may be avoided using the alternative
method of quantization, based on the equivalence theorems  for spectrum changing transformation.
Here the technic of BRST quantization is heavily used.

We shall apply this technique to the Yang-Mills theory
\begin{equation}
L=\frac{1}{8g^2}tr(F_{\mu \nu}F_{\mu \nu})
\label{10}
\end{equation}
This theory is equivalent  to the theory described by the Lagrangian
\begin{eqnarray}
\tilde{L}=\frac{1}{8g^2}tr(F_{\mu \nu}F_{\mu \nu})+(D_\mu \varphi_+)^*(D_\mu\varphi_-)+(D_\mu \varphi_+)(D_\mu\varphi_-)^*\nonumber\\
-[(D_\mu b)^*(D_\mu e)+(D_\mu e)^*(D_\mu b)]
\label{11}
\end{eqnarray}
Here the fields $\varphi_\pm$ commute and $b,e$ are anti commuting elements of Grassman algebra. 
Indeed integrating over $\varphi_\pm, b, e$ in the sector which does not contain these fields in the asymptotic states, we get the usual Yang-Mills theory.
Of course one should prove that the theory is unitary in this sector.
This Lagrangian produces more excitations than the standard Yang-Mills Lagrangian, 
but it also possesses the additional invariance. It is invariant with respect to supertransformations
\begin{equation}
\delta \varphi_-=b\epsilon; \quad \delta e =\varphi_+\epsilon
\label{12}
\end{equation}
Obviously this invariance is preserved if we perform before the shift of the fields $\varphi_-$
\begin{equation}
\varphi_- \rightarrow \varphi_- +m
\label{13}
\end{equation}
This symmetry leads to decoupling of excitations corresponding to the fields
$\varphi_\pm, b, e$.
It allows also to use the gauge $\varphi_-=0$.
This gauge is algebraic and therefore does not produce the Gribov ambiguity, at the same
time it is manifestly Lorentz invariant. 
This gauge does not lead to Gribov ambiguity, and it may be used 
beyond perturbation theory. It goes without saying that one should prove the decoupling of all unphysical 
degrees of freedom. It was done in the papers (\cite{Sl1}, \cite{Sl2},\cite{Sl3}, \cite{Sl4}, \cite{QS}), cited above.

 \centerline{The spectrum:}
Ghost exitations: $\varphi_{\pm},b,e,$ longitudinal and temporal components of
$A_{\mu}^a$\\
Physical exitations: three dimensionally transversal components of the Yang-Mills field.
The supersymmetry of the effective action together with BRST-invariance generates a conserved
nilpotent charge $Q$. Physical states are separated by the condition
\begin{equation}
Q|\psi>_{ph}=0 \label{14}
\end{equation}
 The states separated by this condition describe
 only three dimensionally transversal components of the Yang-Mills field.
There exists an invariant subspace of states annihilated by
$Q$, which has a semidefinite norm (\cite{Sl1},\cite{Sl2}'\cite{Sl3}). For asymptotic space
this condition reduces to
\begin{equation}
Q_0|\phi>_{as}=0
\label{15}
\end{equation}
The structure of the vectors, annihilated by the operators $Q$
may be studied in the following way. It is possible to introduce the operators $K$, satisfying the  conditions
\begin{equation}
[\hat{K}, \hat{Q}]_+=\hat{N}
\label{15a}
\end{equation}
where $\hat{N}$ is the number operator for the ghost particles. 
Any vector which contains at least one ghost particle may be 
presented in the form
\begin{equation}
|\chi>=\frac{[\hat{K}, \hat{Q}]_+}{N}|\chi>=\frac{\hat{K}\hat{Q}}{N}|\chi>+\frac{\hat{Q}\hat{K}}{N}|\chi>
\label{15b}
\end{equation}
The first term is zero for the state vectors annihilated by $\hat{Q}$
The second term has zero norm.
Any vector which satisfies the equation (\ref{14}) has a form
\begin{equation}
|\phi>_{as}=|\phi>_{tr}+|N>
\label{16}
\end{equation}
where $|\phi_{tr}>$ is the physical Yang-Mills vector, having only transversal polarizations, and $|N>$ 
is a zero norm vector,which is proportional to $Q|\lambda>$. Factorizing the space of the vectors, 
satisfying the equation (\ref{15}), with respect to zero norm vectors we  see that the asymptotic space of our model coincides with 
the space of the Yang-Mills theory.

\section{Acknowlegements.}

 This work was supported in part by RBRF under the grant $ofi_m2 13-01-12405$ and the program 
 "Nonlinear dynamics"'.
 
 \end{document}